\pdfoutput=1

\documentclass[11pt,twoside,a4paper,cmspaper,final,collab]{cms-tdr}
\def\svnVersion{41305c8-D}\def\svnDate{2020/02/29}\def\cmsCernNoTag{CERN-EP-2019-211}\def\cmsCernDate{\today}\def\cmsMessage{"Published in the European Physical Journal C as \href{http://dx.doi.org/10.1140/epjc/s10052-020-7834-9}{\doi{10.1140/epjc/s10052-020-7834-9}.}"}

\begin{document}\cmsNoteHeader{HIN-17-005}

\hyphenation{had-ron-i-za-tion}
\hyphenation{cal-or-i-me-ter}
\hyphenation{de-vices}
\RCS$HeadURL$
\RCS$Id$

\newlength\cmsFigWidth
\ifthenelse{\boolean{cms@external}}{\setlength\cmsFigWidth{0.87\textwidth}}{\setlength\cmsFigWidth{0.98\textwidth}}
\ifthenelse{\boolean{cms@external}}{\providecommand{\cmsLeft}{top\xspace}}{\providecommand{\cmsLeft}{left\xspace}}
\ifthenelse{\boolean{cms@external}}{\providecommand{\cmsRight}{bottom\xspace}}{\providecommand{\cmsRight}{right\xspace}}

\newcommand{\PbPb}{\ensuremath{\mathrm{PbPb}}\xspace}

\cmsNoteHeader{HIN-17-005}
\title{Mixed higher-order anisotropic flow and nonlinear response coefficients of charged particles in \PbPb collisions at $\sqrtsNN = 2.76$ and 5.02\TeV}
\date{\today}
\titlerunning{Mixed higer-order anisotropic flow \ldots in \PbPb collisions at $\sqrtsNN = 2.76$ and 5.02\TeV}
\abstract{
Anisotropies in the initial energy density distribution of
the quark-gluon plasma created in high energy heavy ion
collisions lead to anisotropies in the azimuthal distributions
of the final-state particles known as collective anisotropic flow.
Fourier harmonic decomposition is used to quantify these
anisotropies. The higher-order harmonics
can be induced by the same order anisotropies (linear response)
or by the combined influence of several lower order
anisotropies (nonlinear response) in the initial state.
The mixed higher-order anisotropic flow and nonlinear response
coefficients of charged particles are measured as functions of
transverse momentum and centrality in \PbPb collisions at
nucleon-nucleon center-of-mass energies $\sqrtsNN = 2.76$ and 5.02\TeV
with the CMS detector. The results are compared with viscous
hydrodynamic calculations using several different initial conditions,
as well as microscopic transport model calculations. None of the
models provides a simultaneous description of
the mixed higher-order flow harmonics and nonlinear response coefficients.
}

\hypersetup{
pdfauthor={CMS Collaboration},
pdftitle={Mixed higher-order anisotropic flow and nonlinear response coefficients of charged particles in PbPb collisions at sqrt(s[NN]) = 2.76 and 5.02 TeV},
pdfsubject={CMS},
pdfkeywords={CMS, physics, heavy ion, flow, mixing harmonics, nonlinear response coefficients}}

\maketitle

\section{Introduction}
\label{sec:introduction}

The azimuthal anisotropy of particle production
in a heavy ion collision can be characterized by
the Fourier expansion of the particle azimuthal angle
distribution~\cite{Yan:2015jma},
\begin{linenomath}
\begin{equation}
\label{defVn}
\frac{dN}{d\phi}=\frac{N}{2\pi}\sum_{n=-\infty}^{+\infty}V_n e^{-in\phi},
\end{equation}
\end{linenomath}
where $V_n=v_n\exp(in\Psi_n)$ is the
$n$-th complex anisotropic flow coefficient~\cite{Voloshin:1994mz}.
The $v_n$ and $\Psi_n$ are the magnitude and
phase (also known as the $n$-th order symmetry plane angle) of $V_n$, respectively.
Anisotropic flow plays a major role in
probing the properties of the produced medium in heavy ion collisions at the BNL
RHIC~\cite{PHENIX,STAR,PHOBOS,BRAHMS} and CERN
LHC~\cite{Chatrchyan:2012ta,Aamodt:2010pa,ATLAS:2011ah}.
Studies of flow harmonics higher than the 
second order~\cite{Alver:2010gr, Chatrchyan:2013kba, ALICE:2011ab},
flow fluctuations~\cite{Alver:2008zza, Sorensen:2008zk, Alver:2010rt, Ollitrault:2009ie},
the correlation between the magnitude
and phase of different harmonics~\cite{Qiu:2011iv, Adare:2011tg, Niemi:2012aj, Aad:2014fla, Aad:2015lwa, ALICE:2016kpq, Sirunyan:2017uyl, STAR:2018fpo},
and the transverse momentum (\pt) and pseudorapidity $(\eta)$
dependence of symmetry plane angles~\cite{Heinz:2013bua, Khachatryan:2015oea}, have led to a broader and deeper understanding of the
initial conditions~\cite{PHENIX, Busza:2018rrf} and the properties
of the produced hot and dense matter.
There are significant correlations between the
symmetry plane angles of different orders~\cite{Aad:2014fla}, which indicate that
higher-order mixed harmonics can be studied with respect to
multiple lower-order symmetry plane angles.

In hydrodynamical models describing the quark-gluon plasma (QGP)
created in relativistic heavy ion collisions,
anisotropic flow arises from the evolution of the medium in the
presence of an anisotropy in the initial-state energy density,
as characterized by the 
eccentricities $\epsilon_n$~\cite{Alver:2010gr}.
The magnitudes of the second- and third-order harmonic final state coefficients, $v_{2}$ and $v_{3}$, are to a good approximation
linearly proportional to the initial-state anisotropies, $\epsilon_{2}$ and
$\epsilon_{3}$, respectively~\cite{Alver:2010gr, Qiu:2011iv}.
In contrast, $V_4$ and higher harmonics can arise from
initial-state anisotropies in the same-order harmonic (linear response) or can be induced by
lower-order harmonics (nonlinear response)~\cite{Yan:2015jma, Qian:2016fpi, Teaney:2012ke}.
More specifically, these harmonics can be decomposed into linear and nonlinear
response contributions as follows~\cite{Yan:2015jma, Qian:2016fpi}:
\begin{linenomath}
\begin{equation}
\label{decompositionVn}
\begin{split}
  V_4 & = V_{4 L} + \chi_{422} V_2^2,\\
  V_5 & = V_{5 L} + \chi_{523} V_2 V_3,\\
  V_6 & = V_{6 L} + \chi_{624} V_2 V_{4L} + \chi_{633} V_3^2 + \chi_{6222}  V_2^3,\\
  V_7 & = V_{7 L} + \chi_{725} V_2 V_{5L} + \chi_{734} V_3 V_{4L} + \chi_{7223}  V_2^2 V_3,
\end{split}
\end{equation}
\end{linenomath}
where $V_{nL}$ denotes the part of $V_n$ that is not induced
by lower-order harmonics~\cite{Teaney:2012ke, Qian:2017ier, Giacalone:2018wpp}, and
the $\chi$\space are the nonlinear response coefficients.
Each nonlinear response coefficient has its associated mixed harmonic,
which is $V_n$ measured
with respect to the lower-order symmetry plane angle or angles.
The strength of each nonlinear response coefficient determines the
magnitude of its associated mixed harmonic.
The $V_1$ terms are neglected in the decomposition
in Eq.~(\ref{decompositionVn}) because the correlation
between $V_{n}$ and $V_{1}V_{n-1}$ was shown to be
negligible after correcting $V_{1}$ for global
momentum conservation~\cite{Qian:2016fpi}.
This analysis focuses on the terms that only involve the
two largest anisotropic
flow coefficients $V_2$ and $V_3$ on the right-hand side
of Eq.~(\ref{decompositionVn}). 
The procedures used to extract both
mixed-harmonic and nonlinear response coefficients
are given in Section~\ref{sec:analysistechnique}.

It is difficult to use measured $v_2$ and $v_3$ coefficients to
evaluate hydrodynamic theories because these flow observables
have a strong dependence on the initial anisotropies, which
cannot be experimentally determined or tightly constrained.
In contrast, most of the nonlinear response coefficients
are not strongly sensitive to the initial anisotropies, which
largely cancel in the dimensionless
ratios used to determine these
coefficients~\cite{Yan:2015jma, Zhao:2017yhj, Qian:2016fpi, Giacalone:2018wpp}. As
a result, their experimental values can serve as
unique and robust probes of hydrodynamic behavior
of the QGP~\cite{Giacalone:2018wpp}.

Most previous flow measurements focused on $V_n$ (overall flow), \ie, $v_n$ with
respect to $\Psi_n$, which
does not separate the linear and nonlinear parts of Eq.~(\ref{decompositionVn}).
Direct measurements of the mixed higher-order flow
harmonics, $v_4$ and $v_6$ with respect to $\Psi_2$, already
exist at both RHIC~\cite{Adams:2003zg} and LHC~\cite{Chatrchyan:2013kba} energies, but
were performed using the event plane method~\cite{Poskanzer:1998yz}.
This method has been criticized for
yielding an ambiguous measure lying somewhere between the event-averaged
mean value $\left\langle v_n \right\rangle$ and the root-mean-square
value $\sqrt{\left\langle {v_n^2} \right\rangle}$ of the $v_n$ distribution,
depending on the resolution of
the method~\cite{Alver:2008zza, Ollitrault:2009ie, Luzum:2012da}.
This ambiguity can be removed by using the scalar-product
method~\cite{Adler:2002pu, Luzum:2012da},
which always measures the root-mean-square values of $v_n$.
The difference between the two methods is typically a few percent for $v_2$,
$\sim$10\% for $v_3$, and much larger for mixed harmonics~\cite{Luzum:2012da}.

This paper presents the mixed higher-order flow harmonics
and nonlinear response coefficients for $n = 4$, 5, 6, and 7 using the scalar-product method.
These variables are measured in \PbPb collisions
at nucleon-nucleon center-of-mass
energies $\sqrtsNN = 2.76$ and 5.02\TeV, as functions of collision centrality
and charged particle \pt in the
region $\abs{\eta}<0.8$.
To compare the mixed flow harmonics with the overall flow coefficients, the
higher-order flow harmonics with respect to the same-order symmetry
plane, measured using the scalar-product method, are also presented.

\section{The CMS detector}
\label{sec:cmsdetector}

The central feature of the CMS apparatus is a superconducting solenoid of 6\unit{m} internal diameter,
providing a nearly constant magnetic field of 3.8\unit{T}. Within the solenoid volume are a silicon pixel and strip
tracker, a lead tungstate crystal electromagnetic calorimeter, and a brass and scintillator
hadron calorimeter, each composed of a barrel and two endcap sections.
In this analysis, the tracker and the forward hadron (HF) calorimeter
subsystems are of particular importance.
The HF uses steel as an absorber and quartz fibers as the sensitive
material. The two halves of the HF are located 11.2\unit{m} from the
center of the interaction region,
one on each end, and together they provide coverage in the range $3.0 < \abs{\eta} < 5.2$.
These calorimeters are azimuthally subdivided into $20^{\circ}$
modular wedges and further segmented to
form $0.175{\times}0.175$ $(\Delta\eta{\times}\Delta\phi)$ ``towers",
where the angle $\phi$ is in radians.
The silicon tracker measures charged particles within the range $\abs{\eta} < 2.5$.
It consists of 1440 silicon pixel and 15\,148 silicon strip detector modules. For nonisolated
particles of $1 < \pt < 10\GeVc$ and $\abs{\eta} < 1.4$, the track resolutions are
typically 1.5\% in \pt and 25--90 (45--150)\mum in the transverse (longitudinal) impact parameter~\cite{Chatrchyan:2014fea}.
The Beam Pick-up Timing for the eXperiments (BPTX) devices are located around the beam pipe
at a distance of 175\unit{m} from the interaction region on both sides, and are designed to provide
precise information on the LHC bunch structure and timing of the incoming beams.
A more detailed description of the CMS detector, together with a definition of the coordinate system
used and the relevant kinematic variables, can be found in Ref.~\cite{Chatrchyan:2008zzk}.
The Monte Carlo simulation of the particle propagation and detector response is
based on the \GEANTfour~\cite{Agostinelli:2002hh} program.

\section{Event and track selections}
\label{sec:eventandtrackselections}

This analysis is performed using
minimum bias \PbPb data collected with the CMS detector
at $\sqrtsNN = $ 5.02 and 2.76\TeV in 2015 and 2011, corresponding
to integrated luminosities of 13\mubinv and
3.9\mubinv, respectively.
The minimum bias trigger~\cite{Khachatryan:2016bia} used in this analysis requires
coincident signals in the HF calorimeters at both ends of the CMS detector with
total energy deposits above a predefined energy threshold of approximately 1\GeV and
the presence of both colliding bunches in the interaction
region as determined using the BPTX.
By requiring colliding bunches, events due to
noise (\eg, cosmic rays and beam backgrounds) are largely suppressed.
In the offline analysis, events are required to have at
least one reconstructed primary vertex, which is chosen as the
reconstructed vertex with the largest number of associated tracks.
The primary vertex is formed
by two or more associated tracks and is required to have a distance
of less than 15\unit{cm} along the beam axis from the center of the nominal interaction region
and less than 0.15\unit{cm} from the beam position in the transverse plane.
An additional selection of hadronic collisions is applied by
requiring at least three towers, each with
total energy above 3\GeV in each of the two HF calorimeters.
The average number of collisions per bunch crossing is less than
0.001 for the events used in this analysis, with a
pileup fraction less than 0.05\%, which has a negligible
effect on the results.
Events are classified using a centrality variable that is related to
the degree of geometric overlap between the two colliding nuclei. Events with
complete (no) overlap are denoted as centrality 0 (100)\%, where
the number is the fraction of events in a
given class with respect to the total number of inelastic hadronic collisions.
The centrality is determined offline via the sum of the HF energies in each event.
Very central events (centrality approaching 0\%) are characterized
by a large energy deposit in the HF calorimeters.
The results reported in this paper are presented up to 60\% in centrality.
The minimum bias trigger and event selections are fully
efficient in this centrality range.

Track reconstruction~\cite{Chatrchyan:2014fea,Khachatryan:2016odn} is performed in
two iterations to ease the computational load for high-multiplicity central
\PbPb collisions. The first iteration reconstructs tracks from signals
(``hits'') in the silicon pixel and strip detectors compatible with a
trajectory of $\pt>0.9\GeVc$.
The significance of the separation along the beam
axis ($z$) between the track and the primary vertex, $d_z/\sigma(d_z)$, and the significance
of the impact parameter relative to the primary vertex transverse to the beam, $d_{\mathrm{0}}/\sigma(d_{\mathrm{0}})$,
must be less than 2. In addition, the relative uncertainty of the $\pt$ measurement,
$\sigma(\pt)/$\pt, must be less than 5\%, and tracks are required to have
at least 11 out of the possible 14 hits
along their trajectories in the pixel and strip trackers.  To reduce the
number of misidentified tracks, the chi-squared per degree of
freedom, $\chi^2/\mathrm{dof}$, associated with fitting the
track trajectory through the different pixel and strip layers,
must be less than 0.15 times the total number of layers having hits along
the trajectory of the track. The second iteration reconstructs tracks compatible
with a trajectory of $\pt>0.2\GeVc$ using solely the pixel detector. These tracks
are required to have $d_{z}/\sigma(d_{z}) < 6$ and a
fit $\chi^2/\mathrm{dof}$ value less
than 9 times the number of layers with hits along the trajectory
of the track.  In the final analysis,
first iteration tracks with $\pt > 1.0\GeVc$ are combined with pixel-detector-only
tracks that have $0.2 < \pt < 2.4\GeVc$.
After removing duplicates~\cite{Chatrchyan:2012ta},
the merged track collection
has a combined geometric acceptance and efficiency exceeding 60\% for
\pt$\approx 1.0\GeVc$ and $\abs{\eta}<0.8$, as determined
using the \HYDJET event generator~\cite{Lokhtin:2005px}.
When the track \pt is below 1\GeVc,
the acceptance and efficiency steadily drops, reaching approximately 40\%
at $\pt\approx 0.3\GeVc$, which is the lower limit for \pt in this analysis.

\section{Analysis technique}
\label{sec:analysistechnique}

The analysis technique follows the method described in Refs.~\cite{Yan:2015jma, Qian:2016fpi} using detector information from both HF and the tracker.
The notation $V_n=v_n\exp(in\Psi_n)=\left\langle e^{in\phi} \right\rangle$ in
Eq.~(\ref{defVn}) will be replaced by the measured complex flow
vector $Q_n$ with real and imaginary parts defined as
\ifthenelse{\boolean{cms@external}}{
\begin{linenomath}
\begin{multline}
\operatorname{Re}(Q_n) =\\ \frac{1}{\sum{w_j}}\sum\limits_j^M {w_j}\cos \left( {n{\phi_j}} \right) - \left\langle \frac{1}{\sum{w_j}}\sum\limits_j^M {w_j}\cos \left(  {n{\phi_j}} \right) \right\rangle,
\label{eq:qn_cos_recentering}
\end{multline}
\begin{multline}
\operatorname{Im}(Q_n) =\\ \frac{1}{\sum{w_j}}\sum\limits_j^M {w_j}\sin \left( {n{\phi_j}} \right) - \left\langle \frac{1}{\sum{w_j}}\sum\limits_j^M {w_j}\sin \left(  {n{\phi_j}} \right) \right\rangle,
\label{eq:qn_sin_recentering}
\end{multline}
\end{linenomath}
}
{
\begin{linenomath}
\begin{equation}
\operatorname{Re}(Q_n) = \frac{1}{\sum{w_j}}\sum\limits_j^M {w_j}\cos \left( {n{\phi_j}} \right) - \left\langle \frac{1}{\sum{w_j}}\sum\limits_j^M {w_j}\cos \left(  {n{\phi_j}} \right) \right\rangle,
\label{eq:qn_cos_recentering}
\end{equation}
\begin{equation}
\operatorname{Im}(Q_n) = \frac{1}{\sum{w_j}}\sum\limits_j^M {w_j}\sin \left( {n{\phi_j}} \right) - \left\langle \frac{1}{\sum{w_j}}\sum\limits_j^M {w_j}\sin \left(  {n{\phi_j}} \right) \right\rangle,
\label{eq:qn_sin_recentering}
\end{equation}
\end{linenomath}
}
where $M$ represents the number of tracks or HF
towers used for calculating the $Q$ vector, $\phi_j$ is the
azimuthal angle of the $j$-th track or HF tower, and $w_j$ is
a weighting factor equal to transverse energy for HF $Q$ vectors.
To correct for the tracking inefficiency, $w_j = 1/\varepsilon_j$ is the
inverse of the tracking efficiency $\varepsilon_j(\pt, \eta)$ of the $j$-th track.
Unlike the averages over particles in a single event in the definitions
of $Q_n$, the angle brackets in
Eqs.~(\ref{eq:qn_cos_recentering}) and (\ref{eq:qn_sin_recentering}) denote an
average over all the events within a given centrality range.
Subtraction of the event-averaged quantity removes biases due to the detector acceptance.

The mixed higher-order harmonics in each \pt range are extracted
using the scalar-product method as shown
in Eqs.~(\ref{eq:v4psi22_QdiffFinal})--(\ref{eq:v7psi223_QdiffFinal})~\cite{Yan:2015jma},
which describe the various harmonics measured with respect to
symmetry plane angles of different orders.
Equations~(\ref{eq:v4psi22_QdiffFinal})--(\ref{eq:v7psi223_QdiffFinal})
show $v_4$ with respect to the second-order, $v_5$ with respect to
the second- and third-order, $v_6$ with respect to the second-order,
$v_6$ with respect to the third-order, and $v_7$ with respect to the
second- and third-order symmetry plane angles, respectively.
\begin{linenomath}
\begin{equation}
v_{4}\{\Psi_{22}\} \equiv \frac{{\mathrm{Re}}\langle Q_{4}
  Q_{2A}^{*}Q_{2B}^{*}\rangle}{\sqrt{{\mathrm{Re}}\langle Q_{2A}Q_{2A}Q_{2B}^{*}Q_{2B}^{*}  \rangle}}
\label{eq:v4psi22_QdiffFinal}
\end{equation}
\begin{equation}
v_{5}\{\Psi_{23}\} \equiv \frac{{\mathrm{Re}}\langle Q_{5}
  Q_{2A}^{*}Q_{3B}^{*}\rangle}{\sqrt{{\mathrm{Re}}\langle  Q_{2A}Q_{3A}Q_{2B}^{*}Q_{3B}^{*}  \rangle}}
\label{eq:v5psi23_QdiffFinal}
\end{equation}
\begin{equation}
v_{6}\{\Psi_{222}\} \equiv \frac{{\mathrm{Re}}\langle Q_{6}
  Q_{2A}^{*}Q_{2B}^{*}Q_{2B}^{*}\rangle}{\sqrt{{\mathrm{Re}}\langle Q_{2A}Q_{2A}Q_{2A}Q_{2B}^{*}Q_{2B}^{*}Q_{2B}^{*}  \rangle}}
\label{eq:v6psi222_QdiffFinal}
\end{equation}
\begin{equation}
v_{6}\{\Psi_{33}\} \equiv \frac{{\mathrm{Re}}\langle Q_{6}
  Q_{3A}^{*}Q_{3B}^{*}\rangle}{\sqrt{{\mathrm{Re}}\langle Q_{3A}Q_{3A}Q_{3B}^{*}Q_{3B}^{*}  \rangle}}
\label{eq:v6psi33_QdiffFinal}
\end{equation}
\begin{equation}
v_{7}\{\Psi_{223}\} \equiv \frac{{\mathrm{Re}}\langle Q_{7}
  Q_{2A}^{*}Q_{2B}^{*}Q_{3B}^{*}\rangle}{\sqrt{{\mathrm{Re}}\langle Q_{2A}Q_{2A}Q_{3A}Q_{2B}^{*}Q_{2B}^{*}Q_{3B}^{*}  \rangle}}
\label{eq:v7psi223_QdiffFinal}
\end{equation}
\end{linenomath}
Here, $Q_{nA}$ and $Q_{nB}$ are vectors from two different parts
of the detector, specifically the positive and negative sides of HF,
$Q_{n}$ is the vector from charged particles in
each \pt range within $\abs{\eta}<0.8$, and
angle brackets denote the average (weighted by the number of particles)
over all events within a given centrality range.
The minimum $\eta$ gap between
tracks used to find the charged-particle $Q$ vector and towers used for the
HF $Q$ vectors is 2.2 units of $\eta$.

With the assumption that the linear and nonlinear 
terms in Eq.~(\ref{decompositionVn}) are uncorrelated,
the nonlinear response coefficients in each \pt range 
can be expressed as~\cite{Yan:2015jma, Qian:2016fpi},
\begin{linenomath}
\begin{equation}
\chi_{422} = \frac{{\mathrm{Re}}\langle Q_{4}
  Q_{2A}^{*}Q_{2B}^{*}\rangle}{{{\mathrm{Re}}\langle Q_{2}Q_{2}Q_{2A}^{*}Q_{2B}^{*}  \rangle}},
\label{eq:chi422_QdiffFinal}
\end{equation}
\begin{equation}
\chi_{523} = \frac{{\mathrm{Re}}\langle Q_{5}
  Q_{2A}^{*}Q_{3B}^{*}\rangle}{{{\mathrm{Re}}\langle Q_{2}Q_{3}Q_{2A}^{*}Q_{3B}^{*}  \rangle}},
\label{eq:chi523_QdiffFinal}
\end{equation}
\begin{equation}
\chi_{6222} = \frac{{\mathrm{Re}}\langle Q_{6}
  Q_{2A}^{*}Q_{2B}^{*}Q_{2B}^{*}\rangle}{{{\mathrm{Re}}\langle Q_{2}Q_{2}Q_{2}Q_{2A}^{*}Q_{2B}^{*}Q_{2B}^{*}  \rangle}},
\label{eq:chi6222_QdiffFinal}
\end{equation}
\begin{equation}
\chi_{633} = \frac{{\mathrm{Re}}\langle Q_{6}
  Q_{3A}^{*}Q_{3B}^{*}\rangle}{{{\mathrm{Re}}\langle Q_{3}Q_{3}Q_{3A}^{*}Q_{3B}^{*}  \rangle}},
\label{eq:chi633_QdiffFinal}
\end{equation}
\begin{equation}
\chi_{7223} = \frac{{\mathrm{Re}}\langle Q_{7}
  Q_{2A}^{*}Q_{2B}^{*}Q_{3B}^{*}\rangle}{{{\mathrm{Re}}\langle Q_{2}Q_{2}Q_{3}Q_{2A}^{*}Q_{2B}^{*}Q_{3B}^{*}  \rangle}},
\label{eq:chi7223_QdiffFinal}
\end{equation}
\end{linenomath}
where the charged-particle $Q_{n}$ vector enters both the numerator and the denominator.

\section{Systematic uncertainties}
\label{sec:systematicuncertainties}

Six sources of systematic uncertainties are considered in this analysis.
The systematic uncertainty due to vertex position selection is estimated by
comparing the results with events from vertex position ranges
$\abs{v_z}<3$\unit{cm} to $3<\abs{v_z}<15$\unit{cm}.
For both mixed harmonic and nonlinear response coefficients, this uncertainty
is estimated to be 1--3\%, with no dependence on \pt or centrality.
Systematic uncertainty due to track quality requirements are examined by varying the
track selections for $d_z/\sigma(d_z)$ and $d_{\mathrm{0}}/\sigma(d_{\mathrm{0}})$
from 1.5 to 5,
the pixel track $d_{z}/\sigma(d_{z})$ from 5 to 10, and the fit $\chi^2/\mathrm{dof}$ value
from 7 to 18 times the number of layers with hits.
The uncertainty is estimated to be 1--4\% depending on \pt and centrality
for both mixed harmonic and nonlinear response coefficients.

The charged-particle tracking efficiency depends on the efficiency of
detecting different types of charged particles
and the species composition of the set of particles.
Two event generators (\HYDJET~\cite{Lokhtin:2005px}
and \textsc{epos lhc}~\cite{Pierog:2013ria}) with
different particle composition are used to study
the tracking efficiency, and the systematic uncertainty is obtained by
comparing the results using efficiencies from the two generators
mentioned above.
The systematic uncertainty from this source is 3\% for the mixed harmonics and
less than 1\% for the nonlinear response coefficients, with no dependence on \pt or centrality.

The sensitivity of the results to the centrality calibration is evaluated
by varying the trigger and event selection efficiency by $\pm 2$\%.
The resulting uncertainty is estimated to be less than 1\%.
The minimum $\eta$ gap between the correlated charged particles and the $Q$ vectors
in the HF region is changed from 2.2 to 3.2 units of $\eta$ (achieved
by changing the $\eta$ ranges of the HF $Q$ vectors) to
estimate the uncertainty due to short-range correlations from resonance
decays and jets. This study results in a systematic uncertainty
of 1--8\%, depending on both \pt and centrality.
This $\eta$ gap uncertainty also includes a possible physics
effect from the $\eta$-dependent fluctuations
of symmetry plane angles~\cite{Pang:2014pxa,Khachatryan:2015oea}, although
a recent study from the ALICE experiment
indicates that this effect is small for correlations
between symmetry plane angles of different order~\cite{Acharya:2017zfg}.

When the same set of HF towers are used for different $Q$ vectors
in the equations of mixed harmonic and nonlinear response coefficients,
the product of these $Q$ vectors contains self-correlations.
An algorithm for removing the duplicated terms when multiplying two or more
$Q$ vectors, the same as the approach of Ref.~\cite{Bilandzic:2013kga}, is used.
The algorithm only works perfectly when the detector
has fine granularity and there is no merging of HF towers.
Therefore, the difference before and after correcting for this effect is taken
as the systematic uncertainty, yielding 
values which depend on centrality but are always less than 3\%.

The different systematic sources described above are added in quadrature to obtain
the overall systematic uncertainty, which is about 10\% at low \pt and decreases to around
5\% for \pt larger than 1\GeVc. As a function of centrality, the overall systematic uncertainty ranges from 3 to 9\% for different coefficients, with
larger uncertainties for central events.

\section{Results}
\label{sec:results}

\begin{figure*}[htb!]
\centering
\includegraphics[width=0.98\textwidth]{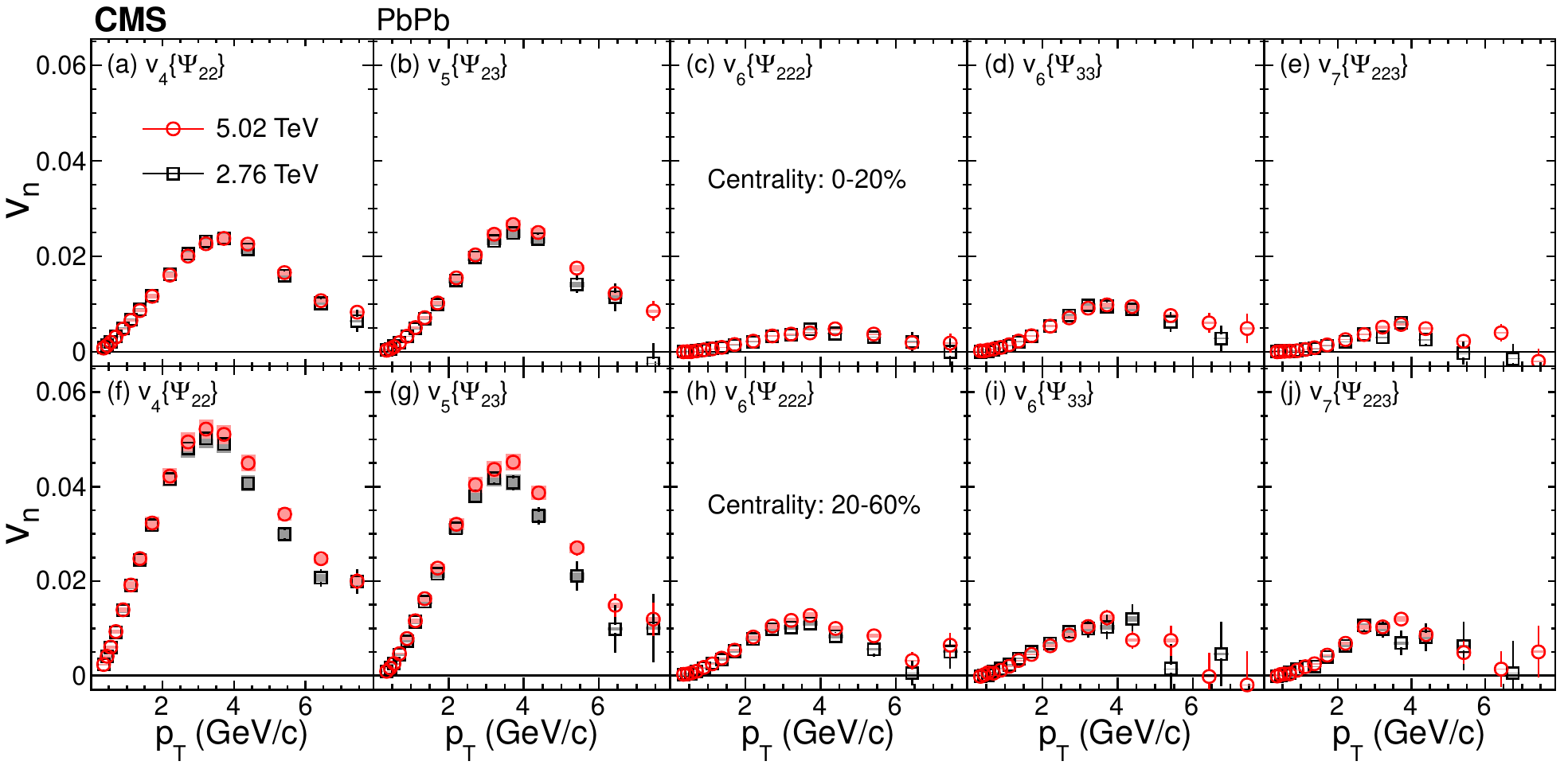}
\caption{Mixed higher-order flow harmonics, $v_4\{\Psi_{22}\}$,
$v_5\{\Psi_{23}\}$, $v_6\{\Psi_{222}\}$, $v_6\{\Psi_{33}\}$, and
$v_7\{\Psi_{223}\}$ from the scalar-product method
at $\sqrtsNN = 2.76$ and 5.02\TeV
as a function of \pt in the 0--20\% (upper row) and
20--60\% (lower row) centrality ranges.
Statistical (bars) and systematic (shaded boxes) uncertainties are shown.}
\label{fig:vnpt_5panels}
\end{figure*}

The measurements in this paper are presented using tracks in the
range of $\abs{\eta} < 0.8$.
Figure~\ref{fig:vnpt_5panels} shows the
mixed higher-order flow harmonics, $v_4\{\Psi_{22}\}$,
$v_5\{\Psi_{23}\}$, $v_6\{\Psi_{222}\}$, $v_6\{\Psi_{33}\}$, and
$v_7\{\Psi_{223}\}$ from the scalar-product method
at $\sqrtsNN = 2.76$ and 5.02\TeV
as a function of \pt in the 0--20\% (upper row) and
20--60\% (lower row) centrality ranges.

\begin{figure*}[htb!]
\centering
\includegraphics[width=0.98\textwidth]{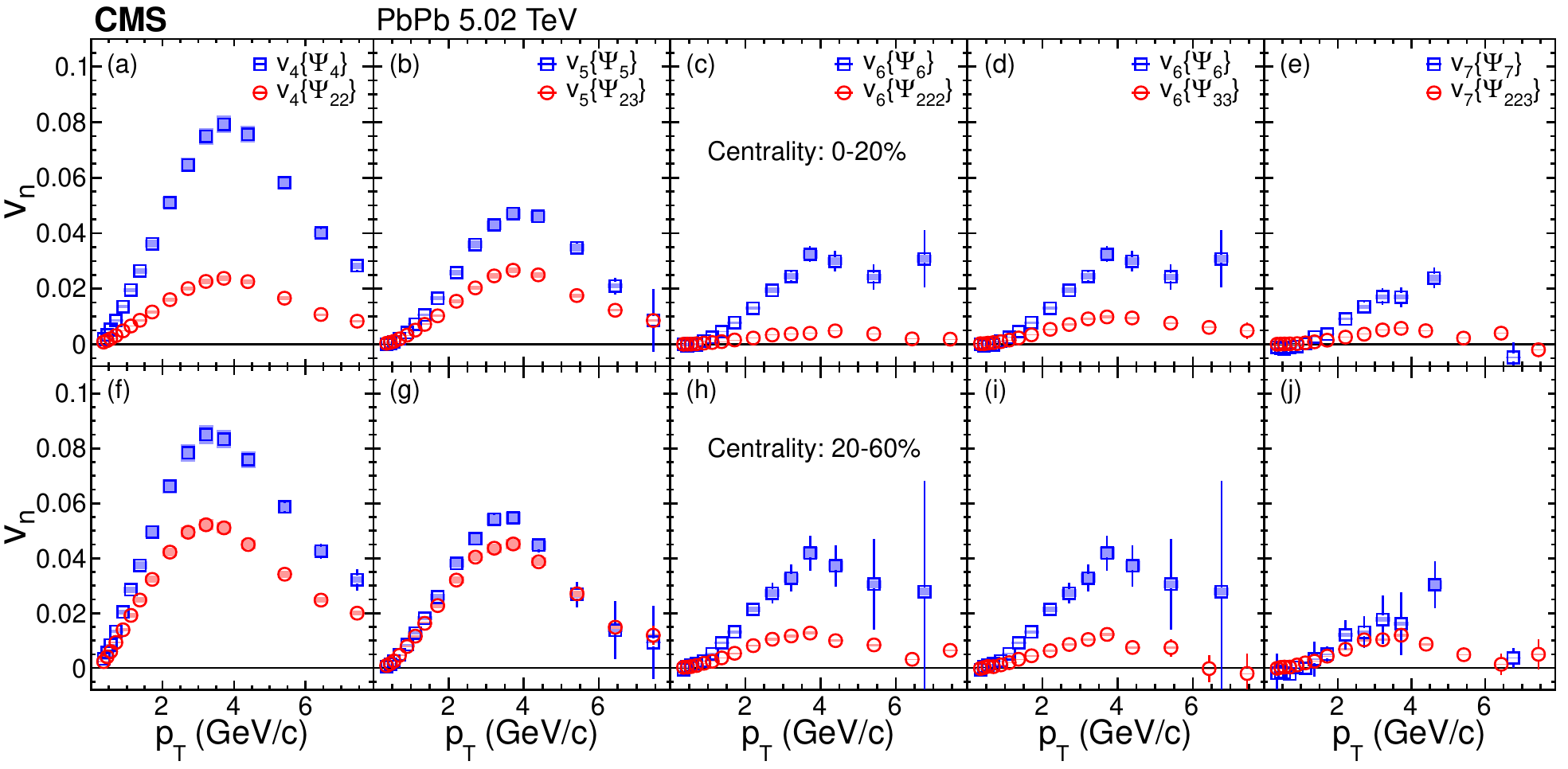}
\caption{Comparison of mixed higher-order flow
harmonics, $v_4\{\Psi_{22}\}$, $v_5\{\Psi_{23}\}$,
$v_6\{\Psi_{222}\}$, $v_6\{\Psi_{33}\}$ and $v_7\{\Psi_{223}\}$
with the corresponding overall flow, $v_4\{\Psi_{4}\}$, $v_5\{\Psi_{5}\}$,
$v_6\{\Psi_{6}\}$, $v_6\{\Psi_{6}\}$ and $v_7\{\Psi_{7}\}$, respectively,
at $\sqrtsNN = 5.02\TeV$ as a function \pt
in the 0--20\% (upper row) and 20--60\% (lower row) centrality ranges.
Statistical (bars) and systematic (shaded boxes) uncertainties are shown.}
\label{fig:vnpt_5panelsComp}
\end{figure*}

It is observed that the shapes of the mixed higher-order flow harmonics
as a function of \pt are qualitatively
similar to the published overall flow harmonics with respect
to $\Psi_n$~\cite{Chatrchyan:2012ta, Chatrchyan:2013kba}, first
increasing at low \pt, reaching a maximum at about 3--4\GeVc, then decreasing at higher \pt.
This may indicate that, for each \pt region, the underlying physics processes that generate
the flow harmonics are the same for the nonlinear and the linear parts.
Similar to previous observation that the overall flow shows a weak energy
dependence from RHIC to LHC energies~\cite{Aamodt:2010pa, Chatrchyan:2012ta},
the mixed harmonics are also found to be consistent between the
two collision energies within the uncertainties, except for
$v_4\{\Psi_{22}\}$ and $v_5\{\Psi_{23}\}$ at \pt larger
than 3\GeVc in the mid-central collisions, with 5.02\TeV results
slightly above 2.76\TeV results.

\begin{figure*}[hbt!]
\centering
\includegraphics[width=0.98\textwidth]{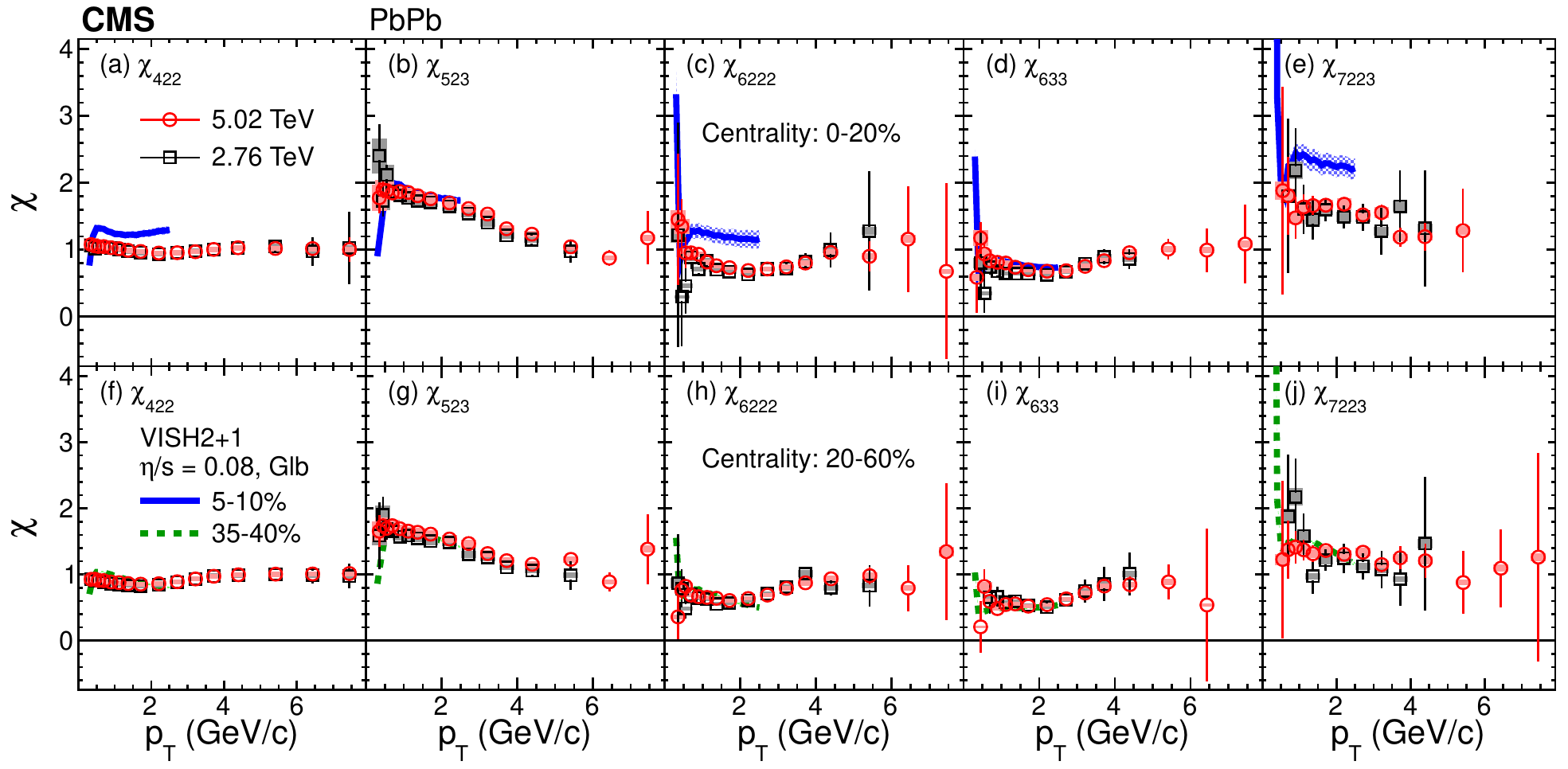}
\caption{Nonlinear response coefficients, $\chi_{422}$,
$\chi_{523}$, $\chi_{6222}$, $\chi_{633}$, and $\chi_{7223}$
from the scalar-product method
at $\sqrtsNN = 2.76$ and 5.02\TeV
as a function of \pt in the 0--20\% (upper row) and
20--60\% (lower row) centrality ranges.
Statistical (bars) and systematic (shaded boxes) uncertainties are shown.
The results are compared with hydrodynamic predictions~\cite{Qian:2017ier}
at $\sqrtsNN = 2.76\TeV$ with $\eta/s = 0.08$ and Glauber initial
conditions in the 5--10\% (blue lines) and
35--40\% (dashed green lines) centrality ranges.
}
\label{fig:chipt_5panelsNew}
\end{figure*}

A direct comparison of the mixed higher-order flow harmonics
and overall flow at 5.02\TeV is presented in
Fig.~\ref{fig:vnpt_5panelsComp} as a function of \pt in the two centrality ranges.
Hydrodynamic models predict that the contribution of the nonlinear response to the overall
flow increases towards peripheral collisions
for $v_4$ and $v_5$~\cite{Teaney:2012ke, Gardim:2011xv, Qiu:2011iv}.
From a comparison of the relative contribution in the
two centrality ranges, the present results
are consistent with these predictions, as well as an estimate by the ATLAS Collaboration
using a two-component fit of the correlation between
flow harmonics~\cite{Aad:2015lwa}, and a
recent study of the nonlinear mode by the ALICE Collaboration~\cite{Acharya:2017zfg}.
By comparing different harmonics, the contribution of the nonlinear response
for $v_5$ is larger than those for the other harmonics in the centrality range 20--60\%.

The nonlinear response coefficients, $\chi_{422}$,
$\chi_{523}$, $\chi_{6222}$, $\chi_{633}$, and $\chi_{7223}$
are presented in Fig.~\ref{fig:chipt_5panelsNew} as a
function of \pt in the two centrality ranges.
It is observed that the odd harmonic coefficients $\chi_{523}$
and $\chi_{7223}$ are larger than those for the even harmonics
for \pt less than 3\GeVc in the two explored centrality ranges.
The values for the even harmonics first decrease slightly
as \pt increases, reach a minimum at \pt about 2\GeVc, and then slowly increase
until appearing to plateau for \pt above 4\GeVc.
The results are compared with viscous hydrodynamic predictions~\cite{Qian:2017ier}
at $\sqrtsNN = 2.76\TeV$ with $\eta/s = 0.08$ (where $\eta/s$ is the
shear viscosity to entropy density ratio of the hydrodynamic medium,
and here $\eta$ denotes shear viscosity rather than
pseudorapidity) and Glauber initial
conditions in two centrality ranges (5--10\% and 35--40\%)
which roughly match those of the data (0--20\% and 20--60\%).
In the model, as \pt increases from 0.3 to 1\GeVc,
the predicted coefficients increase for $n = 4$ and 5,
but decrease and then increase for $n = 6$ and 7, with a
much stronger \pt dependence than the data.
The strong \pt dependence, attributed to the large variance of
the flow angles $\Psi_n$ at small \pt~\cite{Qian:2017ier},
is not observed in data for $n = 4$ and 5.

\begin{figure*}[htb!]
\centering
\includegraphics[width=\cmsFigWidth]{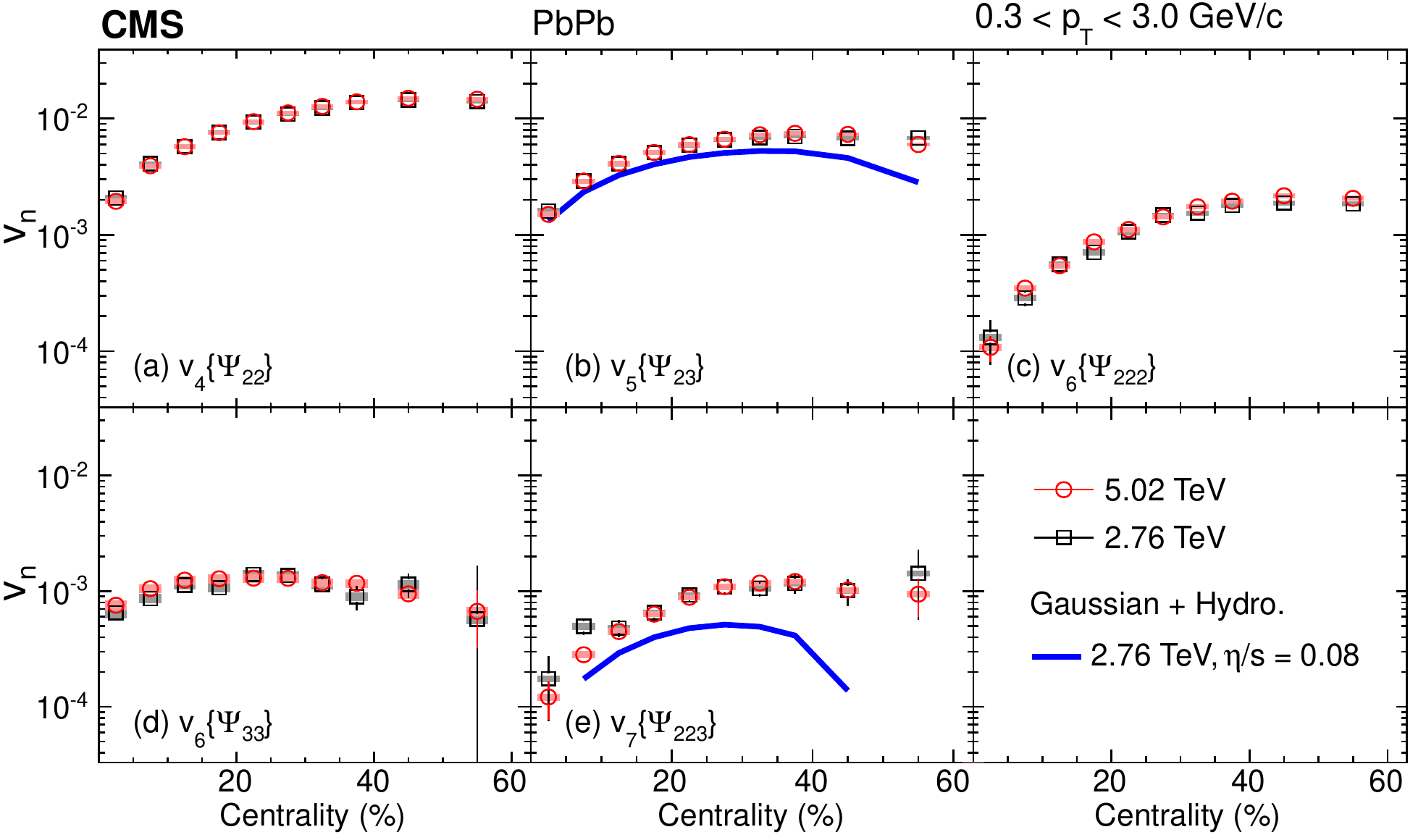}
\caption{Mixed higher-order flow harmonics, $v_4\{\Psi_{22}\}$,
$v_5\{\Psi_{23}\}$, $v_6\{\Psi_{222}\}$, $v_6\{\Psi_{33}\}$, and
$v_7\{\Psi_{223}\}$ from the scalar-product method
at $\sqrtsNN = 2.76$ and 5.02\TeV, as a function of centrality.
Statistical (bars) and systematic (shaded boxes) uncertainties are shown.
Hydrodynamic predictions~\cite{Yan:2015jma} with $\eta/s = 0.08$ (blue lines)
at 2.76\TeV are shown in panel (b) and (e).
}
\label{fig:vncent_theorycomp}
\end{figure*}

Figure~\ref{fig:vncent_theorycomp} shows the
mixed higher-order flow harmonics, $v_4\{\Psi_{22}\}$,
$v_5\{\Psi_{23}\}$, $v_6\{\Psi_{222}\}$, $v_6\{\Psi_{33}\}$, and
$v_7\{\Psi_{223}\}$ from the scalar-product method,
as a function of centrality in the \pt range from 0.3 to 3.0\GeVc.
Hydrodynamic predictions with a deformed symmetric Gaussian density profile as the initial
conditions for $v_5\{\Psi_{23}\}$ and $v_7\{\Psi_{223}\}$~\cite{Yan:2015jma}
at $\sqrtsNN = 2.76\TeV$ are compared with the data.
The model qualitatively describes $v_5\{\Psi_{23}\}$ in the 0--40\% centrality range but
underestimates the result for more peripheral collisions.
For $v_7\{\Psi_{223}\}$, the predicted values are much smaller than the data, especially for
centrality from 35 to 50\%.

\begin{figure*}[htb!]
\centering
\includegraphics[width=\cmsFigWidth]{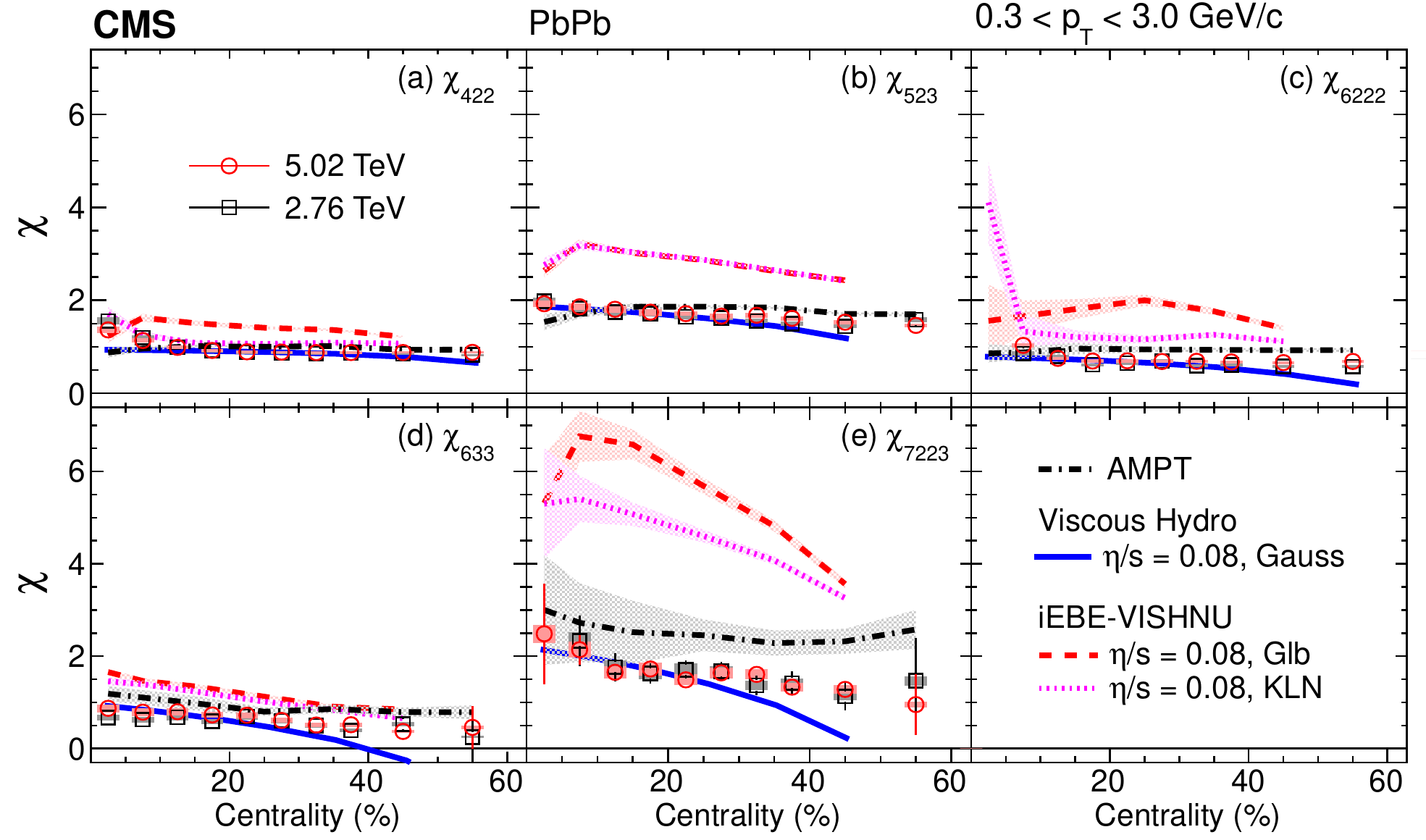}
\caption{Nonlinear response coefficients, $\chi_{422}$,
$\chi_{523}$, $\chi_{6222}$, $\chi_{633}$, and $\chi_{7223}$
from the scalar-product method
at $\sqrtsNN = 2.76$ and 5.02\TeV, as a function of centrality.
Statistical (bars) and systematic (shaded boxes) uncertainties are shown.
The results are compared with predictions at $\sqrtsNN = 2.76\TeV$
from AMPT~\cite{Yan:2015lwn} as well as hydrodynamics
with a deformed symmetric Gaussian density profile as the initial conditions
using $\eta/s = 0.08$ from Ref.~\cite{Yan:2015jma}, and from iEBE-VISHNU
hydrodynamics with both Glauber
and the KLN initial conditions using the same $\eta/s$~\cite{Qian:2016fpi}.
}
\label{fig:chicent_theorycomp}
\end{figure*}

\begin{figure*}[htb!]
\centering
\includegraphics[width=\cmsFigWidth]{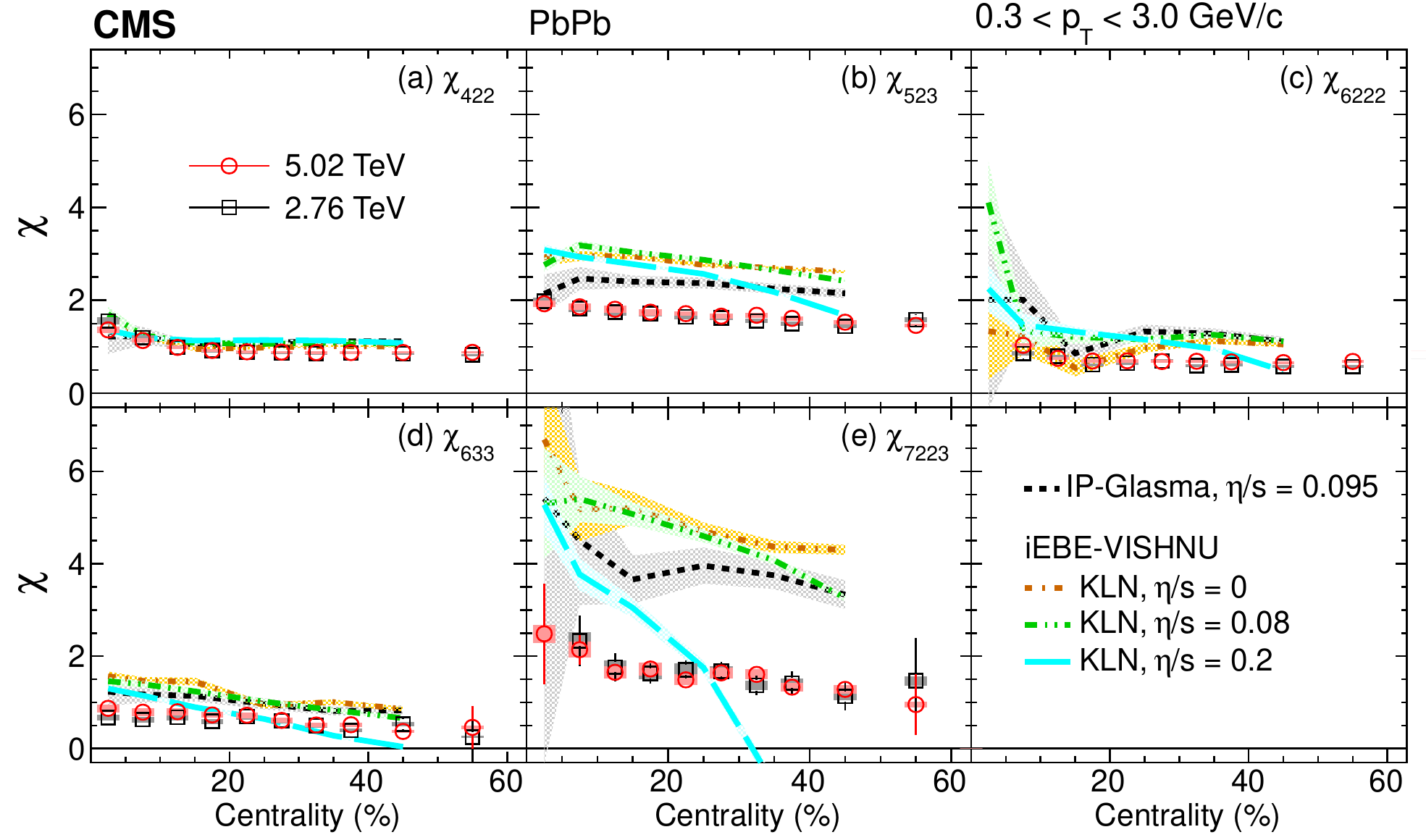}
\caption{The same results as in Fig.~\ref{fig:chicent_theorycomp} but compared with predictions from a hydrodynamics $+$ hadronic cascade
hybrid approach with the IP-Glasma initial conditions
using $\eta/s = 0.095$~\cite{McDonald:2016vlt} at $\sqrtsNN = 5.02\TeV$
and from iEBE-VISHNU hydrodynamics with the KLN
initial conditions using $\eta/s = 0$, 0.08 (the same curve as in
Fig.~\ref{fig:chicent_theorycomp}) and 0.2~\cite{Qian:2016fpi} at $\sqrtsNN = 2.76\TeV$.
}
\label{fig:chicent_theorycompB}
\end{figure*}

The nonlinear response coefficients, $\chi_{422}$,
$\chi_{523}$, $\chi_{6222}$, $\chi_{633}$, and $\chi_{7223}$
are presented in Figs.~\ref{fig:chicent_theorycomp} and~\ref{fig:chicent_theorycompB}, as a
function of centrality in the \pt range from 0.3 to 3.0\GeVc.
The results are compared with predictions at $\sqrtsNN = 2.76\TeV$ from
the microscopic transport model AMPT~\cite{Lin:2004en,Yan:2015lwn},
a macroscopic hydrodynamic model using a deformed symmetric Gaussian density profile as the initial conditions
with $\eta/s = 0.08$~\cite{Yan:2015jma},
and from another hydrodynamic calculation (iEBE-VISHNU) with both Glauber
and Kharzeev-Levin-Nardi (KLN) gluon saturation initial conditions
using the same $\eta/s$~\cite{Qian:2016fpi}.
The model with Gaussian profile initial conditions gives
a better description of the
nonlinear response coefficients compared to other calculations, but it
underestimates the values of $v_7\{\Psi_{223}\}$ for centrality above 30\%,
as shown in Fig.~\ref{fig:vncent_theorycomp}.
In Fig.~\ref{fig:chicent_theorycompB}, the same results are compared with the predictions
from hydrodynamics $+$ hadronic cascade hybrid approach with
the IP-Glasma initial
conditions using $\eta/s = 0.095$~\cite{McDonald:2016vlt} at $\sqrtsNN = 5.02\TeV$
and from iEBE-VISHNU hydrodynamics with the KLN
initial conditions
using $\eta/s = 0$, 0.08 and 0.2~\cite{Qian:2016fpi} at $\sqrtsNN = 2.76\TeV$.
All the calculations describe the $\chi_{422}$ well, but none of them
are successful for $\chi_{523}$ and $\chi_{7223}$.
The  model calculations of $\chi_{7223}$ are
quite different for various initial conditions and
$\eta/s$, which suggests that the first-time measurement of
$\chi_{7223}$ presented in this paper could provide strong constraints on models.

\section{Summary}
\label{sec:summary}

The mixed higher-order flow harmonics and nonlinear response coefficients
of charged particles have been studied as functions of
transverse momentum \pt and centrality
in \PbPb collisions at $\sqrtsNN = 2.76$ and 5.02\TeV using the CMS detector.
The measurements use the scalar-product method, covering a \pt range
from 0.3 to 8.0\GeVc, pseudorapidity $\abs{\eta}<0.8$, and a centrality
range of 0--60\%. The mixed higher-order flow harmonics, $v_4\{\Psi_{22}\}$,
$v_5\{\Psi_{23}\}$, $v_6\{\Psi_{222}\}$, $v_6\{\Psi_{33}\}$, and $v_7\{\Psi_{223}\}$
all have a qualitatively similar \pt dependence,
first increasing at low \pt, reaching a maximum at
about 3--4\GeVc, and then decreasing at higher \pt.
As a comparison, the overall $v_n$ harmonics ($n = 4$--7)
with respect to their own symmetry planes are measured
in the same \pt, $\eta$, and centrality ranges.
The relative contribution of the nonlinear part for $v_5$ is larger than
for other harmonics in the centrality range 20--60\%.
In addition, the nonlinear response coefficients of the odd harmonics are
observed to be larger than those of even harmonics for \pt less than 3\GeVc.
At \pt less than 1\GeVc, a viscous hydrodynamic calculation with
Glauber initial conditions and shear viscosity to entropy density
ratio $\eta/s = 0.08$ predicts a much stronger \pt dependence
for the nonlinear response coefficients.
The coefficients, including the first-time measurement
of $\chi_{7223}$, as a function of centrality, are compared with AMPT
and hydrodynamic predictions using different $\eta/s$ and initial conditions.
Compared to the data, none of the models provides a simultaneous
description of the mixed higher-order flow harmonics and
nonlinear response coefficients.
Therefore, these results can constrain both initial
conditions and transport properties of the produced medium.

\begin{acknowledgments}
We congratulate our colleagues in the CERN accelerator departments for the excellent performance of the LHC and thank the technical and administrative staffs at CERN and at other CMS institutes for their contributions to the success of the CMS effort. In addition, we gratefully acknowledge the computing centers and personnel of the Worldwide LHC Computing Grid for delivering so effectively the computing infrastructure essential to our analyses. Finally, we acknowledge the enduring support for the construction and operation of the LHC and the CMS detector provided by the following funding agencies: BMBWF and FWF (Austria); FNRS and FWO (Belgium); CNPq, CAPES, FAPERJ, FAPERGS, and FAPESP (Brazil); MES (Bulgaria); CERN; CAS, MoST, and NSFC (China); COLCIENCIAS (Colombia); MSES and CSF (Croatia); RPF (Cyprus); SENESCYT (Ecuador); MoER, ERC IUT, PUT and ERDF (Estonia); Academy of Finland, MEC, and HIP (Finland); CEA and CNRS/IN2P3 (France); BMBF, DFG, and HGF (Germany); GSRT (Greece); NKFIA (Hungary); DAE and DST (India); IPM (Iran); SFI (Ireland); INFN (Italy); MSIP and NRF (Republic of Korea); MES (Latvia); LAS (Lithuania); MOE and UM (Malaysia); BUAP, CINVESTAV, CONACYT, LNS, SEP, and UASLP-FAI (Mexico); MOS (Montenegro); MBIE (New Zealand); PAEC (Pakistan); MSHE and NSC (Poland); FCT (Portugal); JINR (Dubna); MON, RosAtom, RAS, RFBR, and NRC KI (Russia); MESTD (Serbia); SEIDI, CPAN, PCTI, and FEDER (Spain); MOSTR (Sri Lanka); Swiss Funding Agencies (Switzerland); MST (Taipei); ThEPCenter, IPST, STAR, and NSTDA (Thailand); TUBITAK and TAEK (Turkey); NASU (Ukraine); STFC (United Kingdom); DOE and NSF (USA).

\hyphenation{Rachada-pisek} Individuals have received support from the Marie-Curie program and the European Research Council and Horizon 2020 Grant, contract Nos.\ 675440, 752730, and 765710 (European Union); the Leventis Foundation; the A.P.\ Sloan Foundation; the Alexander von Humboldt Foundation; the Belgian Federal Science Policy Office; the Fonds pour la Formation \`a la Recherche dans l'Industrie et dans l'Agriculture (FRIA-Belgium); the Agentschap voor Innovatie door Wetenschap en Technologie (IWT-Belgium); the F.R.S.-FNRS and FWO (Belgium) under the ``Excellence of Science -- EOS" -- be.h project n.\ 30820817; the Beijing Municipal Science \& Technology Commission, No. Z181100004218003; the Ministry of Education, Youth and Sports (MEYS) of the Czech Republic; the Lend\"ulet (``Momentum") Program and the J\'anos Bolyai Research Scholarship of the Hungarian Academy of Sciences, the New National Excellence Program \'UNKP, the NKFIA research grants 123842, 123959, 124845, 124850, 125105, 128713, 128786, and 129058 (Hungary); the Council of Science and Industrial Research, India; the HOMING PLUS program of the Foundation for Polish Science, cofinanced from European Union, Regional Development Fund, the Mobility Plus program of the Ministry of Science and Higher Education, the National Science Center (Poland), contracts Harmonia 2014/14/M/ST2/00428, Opus 2014/13/B/ST2/02543, 2014/15/B/ST2/03998, and 2015/19/B/ST2/02861, Sonata-bis 2012/07/E/ST2/01406; the National Priorities Research Program by Qatar National Research Fund; the Ministry of Science and Education, grant no. 3.2989.2017 (Russia); the Programa Estatal de Fomento de la Investigaci{\'o}n Cient{\'i}fica y T{\'e}cnica de Excelencia Mar\'{\i}a de Maeztu, grant MDM-2015-0509 and the Programa Severo Ochoa del Principado de Asturias; the Thalis and Aristeia programs cofinanced by EU-ESF and the Greek NSRF; the Rachadapisek Sompot Fund for Postdoctoral Fellowship, Chulalongkorn University and the Chulalongkorn Academic into Its 2nd Century Project Advancement Project (Thailand); the Nvidia Corporation; the Welch Foundation, contract C-1845; and the Weston Havens Foundation (USA).
\end{acknowledgments}

\bibliography{auto_generated}

\providecommand{\href}[2]{#2}\begingroup\raggedright\begin{thebibliography}{10}%
\makeatletter
\providecommand{\hrefCMSnoop }[0]{\@secondoftwo}%
\makeatother
\providecommand{\doi}{\texttt{doi:}\begingroup \urlstyle{tt}\Url}

\bibitem{Yan:2015jma}
\hrefCMSnoop {}{L.~Yan and J.-Y. Ollitrault, ``{$\nu_4, \nu_5, \nu_6, \nu_7$}:
  nonlinear hydrodynamic response versus {LHC} data'',} \textit{ Phys. Lett. B}
  \textbf{ 744} (2015) 82,
  \href{http://dx.doi.org/10.1016/j.physletb.2015.03.040}{\doi{10.1016/j.physletb.2015.03.040}},
\href{http://www.arXiv.org/abs/1502.02502}{\texttt{arXiv:1502.02502}}.
%%CITATION = ARXIV:1502.02502;%%.

\bibitem{Voloshin:1994mz}
\hrefCMSnoop {}{S.~Voloshin and Y.~Zhang, ``Flow study in relativistic nuclear
  collisions by {Fourier} expansion of azimuthal particle distributions'',}
  \textit{ Z. Phys. C} \textbf{ 70} (1996) 665,
  \href{http://dx.doi.org/10.1007/s002880050141}{\doi{10.1007/s002880050141}},
\href{http://www.arXiv.org/abs/hep-ph/9407282}{\texttt{arXiv:hep-ph/9407282}}.
%%CITATION = HEP-PH/9407282;%%.

\bibitem{PHENIX}
\hrefCMSnoop {}{{PHENIX} Collaboration, ``Formation of dense partonic matter in
  relativistic nucleus-nucleus collisions at {RHIC}: Experimental evaluation by
  the {PHENIX} collaboration'',} \textit{ Nucl. Phys. A} \textbf{ 757} (2005)
  184,
  \href{http://dx.doi.org/10.1016/j.nuclphysa.2005.03.086}{\doi{10.1016/j.nuclphysa.2005.03.086}},
\href{http://www.arXiv.org/abs/nucl-ex/0410003}{\texttt{arXiv:nucl-ex/0410003}}.
%%CITATION = NUCL-EX/0410003;%%.

\bibitem{STAR}
\hrefCMSnoop {}{{STAR} Collaboration, ``Experimental and theoretical challenges
  in the search for the quark gluon plasma: The {STAR} collaboration's critical
  assessment of the evidence from {RHIC} collisions'',} \textit{ Nucl. Phys. A}
  \textbf{ 757} (2005) 102,
  \href{http://dx.doi.org/10.1016/j.nuclphysa.2005.03.085}{\doi{10.1016/j.nuclphysa.2005.03.085}},
\href{http://www.arXiv.org/abs/nucl-ex/0501009}{\texttt{arXiv:nucl-ex/0501009}}.
%%CITATION = NUCL-EX/0501009;%%.

\bibitem{PHOBOS}
\hrefCMSnoop {}{{PHOBOS} Collaboration, ``The {PHOBOS} perspective on
  discoveries at {RHIC}'',} \textit{ Nucl. Phys. A} \textbf{ 757} (2005) 28,
  \href{http://dx.doi.org/10.1016/j.nuclphysa.2005.03.084}{\doi{10.1016/j.nuclphysa.2005.03.084}},
\href{http://www.arXiv.org/abs/nucl-ex/0410022}{\texttt{arXiv:nucl-ex/0410022}}.
%%CITATION = NUCL-EX/0410022;%%.

\bibitem{BRAHMS}
\hrefCMSnoop {}{{BRAHMS} Collaboration, ``{Quark gluon plasma and color glass
  condensate at {RHIC}? The perspective from the {BRAHMS} experiment}'',}
  \textit{ Nucl. Phys. A} \textbf{ 757} (2005) 1,
  \href{http://dx.doi.org/10.1016/j.nuclphysa.2005.02.130}{\doi{10.1016/j.nuclphysa.2005.02.130}},
\href{http://www.arXiv.org/abs/nucl-ex/0410020}{\texttt{arXiv:nucl-ex/0410020}}.
%%CITATION = NUCL-EX/0410020;%%.

\bibitem{Chatrchyan:2012ta}
\hrefCMSnoop {}{{CMS Collaboration}, ``Measurement of the elliptic anisotropy
  of charged particles produced in {PbPb} collisions at {$\sqrtsNN =
  2.76\TeV$}'',} \textit{ Phys. Rev. C} \textbf{ 87} (2013) 014902,
  \href{http://dx.doi.org/10.1103/PhysRevC.87.014902}{\doi{10.1103/PhysRevC.87.014902}},
\href{http://www.arXiv.org/abs/1204.1409}{\texttt{arXiv:1204.1409}}.
%%CITATION = ARXIV:1204.1409;%%.

\bibitem{Aamodt:2010pa}
\hrefCMSnoop {}{{ALICE Collaboration}, ``Elliptic flow of charged particles in
  {Pb-Pb} collisions at 2.76 {\TeV}'',} \textit{ Phys. Rev. Lett.} \textbf{
  105} (2010) 252302,
  \href{http://dx.doi.org/10.1103/PhysRevLett.105.252302}{\doi{10.1103/PhysRevLett.105.252302}},
\href{http://www.arXiv.org/abs/1011.3914}{\texttt{arXiv:1011.3914}}.
%%CITATION = ARXIV:1011.3914;%%.

\bibitem{ATLAS:2011ah}
\hrefCMSnoop {}{{ATLAS Collaboration}, ``Measurement of the pseudorapidity and
  transverse momentum dependence of the elliptic flow of charged particles in
  lead-lead collisions at {$\sqrtsNN=2.76\TeV$} with the {ATLAS} detector'',}
  \textit{ Phys. Lett. B} \textbf{ 707} (2012) 330,
  \href{http://dx.doi.org/10.1016/j.physletb.2011.12.056}{\doi{10.1016/j.physletb.2011.12.056}},
\href{http://www.arXiv.org/abs/1108.6018}{\texttt{arXiv:1108.6018}}.
%%CITATION = ARXIV:1108.6018;%%.

\bibitem{Alver:2010gr}
\hrefCMSnoop {}{B.~Alver and G.~Roland, ``Collision geometry fluctuations and
  triangular flow in heavy-ion collisions'',} \textit{ Phys. Rev. C} \textbf{
  81} (2010) 054905,
  \href{http://dx.doi.org/10.1103/PhysRevC.81.054905}{\doi{10.1103/PhysRevC.81.054905}},
  \href{http://www.arXiv.org/abs/1003.0194}{\texttt{arXiv:1003.0194}}.
[Erratum: \DOI{10.1103/PhysRevC.82.039903}].
%%CITATION = ARXIV:1003.0194;%%.

\bibitem{Chatrchyan:2013kba}
\hrefCMSnoop {}{{CMS Collaboration}, ``Measurement of higher-order harmonic
  azimuthal anisotropy in {PbPb} collisions at {$\sqrtsNN = 2.76\TeV$}'',}
  \textit{ Phys. Rev. C} \textbf{ 89} (2014) 044906,
  \href{http://dx.doi.org/10.1103/PhysRevC.89.044906}{\doi{10.1103/PhysRevC.89.044906}},
\href{http://www.arXiv.org/abs/1310.8651}{\texttt{arXiv:1310.8651}}.
%%CITATION = ARXIV:1310.8651;%%.

\bibitem{ALICE:2011ab}
\hrefCMSnoop {}{{ALICE Collaboration}, ``Higher harmonic anisotropic flow
  measurements of charged particles in {Pb-Pb} collisions at
  {$\sqrtsNN=2.76\TeV$}'',} \textit{ Phys. Rev. Lett.} \textbf{ 107} (2011)
  032301,
  \href{http://dx.doi.org/10.1103/PhysRevLett.107.032301}{\doi{10.1103/PhysRevLett.107.032301}},
\href{http://www.arXiv.org/abs/1105.3865}{\texttt{arXiv:1105.3865}}.
%%CITATION = ARXIV:1105.3865;%%.

\bibitem{Alver:2008zza}
\hrefCMSnoop {}{B.~Alver {et~al.}, ``Importance of correlations and
  fluctuations on the initial source eccentricity in high-energy
  nucleus-nucleus collisions'',} \textit{ Phys. Rev. C} \textbf{ 77} (2008)
  014906,
  \href{http://dx.doi.org/10.1103/PhysRevC.77.014906}{\doi{10.1103/PhysRevC.77.014906}},
\href{http://www.arXiv.org/abs/0711.3724}{\texttt{arXiv:0711.3724}}.
%%CITATION = ARXIV:0711.3724;%%.

\bibitem{Sorensen:2008zk}
\hrefCMSnoop {}{{STAR} Collaboration, ``Elliptic flow fluctuations in {Au+Au}
  collisions at {$\sqrtsNN = 200\GeV$}'',} \textit{ J. Phys. G} \textbf{ 35}
  (2008) 104102,
  \href{http://dx.doi.org/10.1088/0954-3899/35/10/104102}{\doi{10.1088/0954-3899/35/10/104102}},
\href{http://www.arXiv.org/abs/0808.0356}{\texttt{arXiv:0808.0356}}.
%%CITATION = ARXIV:0808.0356;%%.

\bibitem{Alver:2010rt}
\hrefCMSnoop {}{{PHOBOS} Collaboration, ``Non-flow correlations and elliptic
  flow fluctuations in gold-gold collisions at {$\sqrtsNN = 200\GeV$}'',}
  \textit{ Phys. Rev. C} \textbf{ 81} (2010) 034915,
  \href{http://dx.doi.org/10.1103/PhysRevC.81.034915}{\doi{10.1103/PhysRevC.81.034915}},
\href{http://www.arXiv.org/abs/1002.0534}{\texttt{arXiv:1002.0534}}.
%%CITATION = ARXIV:1002.0534;%%.

\bibitem{Ollitrault:2009ie}
\hrefCMSnoop {}{J.-Y. Ollitrault, A.~M. Poskanzer, and S.~A. Voloshin, ``Effect
  of flow fluctuations and nonflow on elliptic flow methods'',} \textit{ Phys.
  Rev. C} \textbf{ 80} (2009) 014904,
  \href{http://dx.doi.org/10.1103/PhysRevC.80.014904}{\doi{10.1103/PhysRevC.80.014904}},
\href{http://www.arXiv.org/abs/0904.2315}{\texttt{arXiv:0904.2315}}.
%%CITATION = ARXIV:0904.2315;%%.

\bibitem{Qiu:2011iv}
\hrefCMSnoop {}{Z.~Qiu and U.~W. Heinz, ``Event-by-event shape and flow
  fluctuations of relativistic heavy-ion collision fireballs'',} \textit{ Phys.
  Rev. C} \textbf{ 84} (2011) 024911,
  \href{http://dx.doi.org/10.1103/PhysRevC.84.024911}{\doi{10.1103/PhysRevC.84.024911}},
\href{http://www.arXiv.org/abs/1104.0650}{\texttt{arXiv:1104.0650}}.
%%CITATION = ARXIV:1104.0650;%%.

\bibitem{Adare:2011tg}
\hrefCMSnoop {}{{PHENIX} Collaboration, ``Measurements of higher-order flow
  harmonics in {Au+Au} collisions at {$\sqrtsNN = 200\GeV$}'',} \textit{ Phys.
  Rev. Lett.} \textbf{ 107} (2011) 252301,
  \href{http://dx.doi.org/10.1103/PhysRevLett.107.252301}{\doi{10.1103/PhysRevLett.107.252301}},
\href{http://www.arXiv.org/abs/1105.3928}{\texttt{arXiv:1105.3928}}.
%%CITATION = ARXIV:1105.3928;%%.

\bibitem{Niemi:2012aj}
\hrefCMSnoop {}{H.~Niemi, G.~S. Denicol, H.~Holopainen, and P.~Huovinen,
  ``Event-by-event distributions of azimuthal asymmetries in ultrarelativistic
  heavy-ion collisions'',} \textit{ Phys. Rev. C} \textbf{ 87} (2013) 054901,
  \href{http://dx.doi.org/10.1103/PhysRevC.87.054901}{\doi{10.1103/PhysRevC.87.054901}},
\href{http://www.arXiv.org/abs/1212.1008}{\texttt{arXiv:1212.1008}}.
%%CITATION = ARXIV:1212.1008;%%.

\bibitem{Aad:2014fla}
\hrefCMSnoop {}{{ATLAS Collaboration}, ``Measurement of event-plane
  correlations in {$\sqrtsNN = 2.76\TeV$} lead-lead collisions with the {ATLAS}
  detector'',} \textit{ Phys. Rev. C} \textbf{ 90} (2014) 024905,
  \href{http://dx.doi.org/10.1103/PhysRevC.90.024905}{\doi{10.1103/PhysRevC.90.024905}},
\href{http://www.arXiv.org/abs/1403.0489}{\texttt{arXiv:1403.0489}}.
%%CITATION = ARXIV:1403.0489;%%.

\bibitem{Aad:2015lwa}
\hrefCMSnoop {}{{ATLAS Collaboration}, ``Measurement of the correlation between
  flow harmonics of different order in lead-lead collisions at {$\sqrtsNN =
  2.76\TeV$} with the {ATLAS} detector'',} \textit{ Phys. Rev. C} \textbf{ 92}
  (2015) 034903,
  \href{http://dx.doi.org/10.1103/PhysRevC.92.034903}{\doi{10.1103/PhysRevC.92.034903}},
\href{http://www.arXiv.org/abs/1504.01289}{\texttt{arXiv:1504.01289}}.
%%CITATION = ARXIV:1504.01289;%%.

\bibitem{ALICE:2016kpq}
\hrefCMSnoop {}{{ALICE Collaboration}, ``Correlated event-by-event fluctuations
  of flow harmonics in {Pb-Pb} collisions at {$\sqrtsNN=2.76\TeV$}'',} \textit{
  Phys. Rev. Lett.} \textbf{ 117} (2016) 182301,
  \href{http://dx.doi.org/10.1103/PhysRevLett.117.182301}{\doi{10.1103/PhysRevLett.117.182301}},
\href{http://www.arXiv.org/abs/1604.07663}{\texttt{arXiv:1604.07663}}.
%%CITATION = ARXIV:1604.07663;%%.

\bibitem{Sirunyan:2017uyl}
\hrefCMSnoop {}{{CMS Collaboration}, ``Observation of correlated azimuthal
  anisotropy {Fourier} harmonics in {pp} and {p+Pb} collisions at the {LHC}'',}
  \textit{ Phys. Rev. Lett.} \textbf{ 120} (2018) 092301,
  \href{http://dx.doi.org/10.1103/PhysRevLett.120.092301}{\doi{10.1103/PhysRevLett.120.092301}},
\href{http://www.arXiv.org/abs/1709.09189}{\texttt{arXiv:1709.09189}}.
%%CITATION = ARXIV:1709.09189;%%.

\bibitem{STAR:2018fpo}
\hrefCMSnoop {}{{STAR} Collaboration, ``Correlation measurements between flow
  harmonics in {Au+Au} collisions at {RHIC}'',} \textit{ Phys. Lett. B}
  \textbf{ 783} (2018) 459,
  \href{http://dx.doi.org/10.1016/j.physletb.2018.05.076}{\doi{10.1016/j.physletb.2018.05.076}},
\href{http://www.arXiv.org/abs/1803.03876}{\texttt{arXiv:1803.03876}}.
%%CITATION = ARXIV:1803.03876;%%.

\bibitem{Heinz:2013bua}
\hrefCMSnoop {}{U.~Heinz, Z.~Qiu, and C.~Shen, ``Fluctuating flow angles and
  anisotropic flow measurements'',} \textit{ Phys. Rev. C} \textbf{ 87} (2013)
  034913,
  \href{http://dx.doi.org/10.1103/PhysRevC.87.034913}{\doi{10.1103/PhysRevC.87.034913}},
\href{http://www.arXiv.org/abs/1302.3535}{\texttt{arXiv:1302.3535}}.
%%CITATION = ARXIV:1302.3535;%%.

\bibitem{Khachatryan:2015oea}
\hrefCMSnoop {}{{CMS Collaboration}, ``Evidence for transverse momentum and
  pseudorapidity dependent event plane fluctuations in {PbPb} and {pPb}
  collisions'',} \textit{ Phys. Rev. C} \textbf{ 92} (2015) 034911,
  \href{http://dx.doi.org/10.1103/PhysRevC.92.034911}{\doi{10.1103/PhysRevC.92.034911}},
\href{http://www.arXiv.org/abs/1503.01692}{\texttt{arXiv:1503.01692}}.
%%CITATION = ARXIV:1503.01692;%%.

\bibitem{Busza:2018rrf}
\hrefCMSnoop {}{W.~Busza, K.~Rajagopal, and W.~van~der Schee, ``Heavy ion
  collisions: The big picture, and the big questions'',} \textit{ Ann. Rev.
  Nucl. Part. Sci.} \textbf{ 68} (2018) 339,
  \href{http://dx.doi.org/10.1146/annurev-nucl-101917-020852}{\doi{10.1146/annurev-nucl-101917-020852}},
\href{http://www.arXiv.org/abs/1802.04801}{\texttt{arXiv:1802.04801}}.
%%CITATION = ARXIV:1802.04801;%%.

\bibitem{Qian:2016fpi}
\hrefCMSnoop {}{J.~Qian, U.~W. Heinz, and J.~Liu, ``Mode-coupling effects in
  anisotropic flow in heavy-ion collisions'',} \textit{ Phys. Rev. C} \textbf{
  93} (2016) 064901,
  \href{http://dx.doi.org/10.1103/PhysRevC.93.064901}{\doi{10.1103/PhysRevC.93.064901}},
\href{http://www.arXiv.org/abs/1602.02813}{\texttt{arXiv:1602.02813}}.
%%CITATION = ARXIV:1602.02813;%%.

\bibitem{Teaney:2012ke}
\hrefCMSnoop {}{D.~Teaney and L.~Yan, ``Non linearities in the harmonic
  spectrum of heavy ion collisions with ideal and viscous hydrodynamics'',}
  \textit{ Phys. Rev. C} \textbf{ 86} (2012) 044908,
  \href{http://dx.doi.org/10.1103/PhysRevC.86.044908}{\doi{10.1103/PhysRevC.86.044908}},
\href{http://www.arXiv.org/abs/1206.1905}{\texttt{arXiv:1206.1905}}.
%%CITATION = ARXIV:1206.1905;%%.

\bibitem{Qian:2017ier}
\hrefCMSnoop {}{J.~Qian, U.~Heinz, R.~He, and L.~Huo, ``Differential flow
  correlations in relativistic heavy-ion collisions'',} \textit{ Phys. Rev. C}
  \textbf{ 95} (2017) 054908,
  \href{http://dx.doi.org/10.1103/PhysRevC.95.054908}{\doi{10.1103/PhysRevC.95.054908}},
\href{http://www.arXiv.org/abs/1703.04077}{\texttt{arXiv:1703.04077}}.
%%CITATION = ARXIV:1703.04077;%%.

\bibitem{Giacalone:2018wpp}
\hrefCMSnoop {}{G.~Giacalone, L.~Yan, and J.-Y. Ollitrault, ``Nonlinear
  coupling of flow harmonics: Hexagonal flow and beyond'',} \textit{ Phys. Rev.
  C} \textbf{ 97} (2018) 054905,
  \href{http://dx.doi.org/10.1103/PhysRevC.97.054905}{\doi{10.1103/PhysRevC.97.054905}},
\href{http://www.arXiv.org/abs/1803.00253}{\texttt{arXiv:1803.00253}}.
%%CITATION = ARXIV:1803.00253;%%.

\bibitem{Zhao:2017yhj}
\hrefCMSnoop {}{W.~Zhao, H.-j. Xu, and H.~Song, ``Collective flow in {2.76 A
  \TeV} and {5.02 A \TeV} {Pb+Pb} collisions'',} \textit{ Eur. Phys. J. C}
  \textbf{ 77} (2017) 645,
  \href{http://dx.doi.org/10.1140/epjc/s10052-017-5186-x}{\doi{10.1140/epjc/s10052-017-5186-x}},
\href{http://www.arXiv.org/abs/1703.10792}{\texttt{arXiv:1703.10792}}.
%%CITATION = ARXIV:1703.10792;%%.

\bibitem{Adams:2003zg}
\hrefCMSnoop {}{{STAR} Collaboration, ``Azimuthal anisotropy at {RHIC}: The
  first and fourth harmonics'',} \textit{ Phys. Rev. Lett.} \textbf{ 92} (2004)
  062301,
  \href{http://dx.doi.org/10.1103/PhysRevLett.92.062301}{\doi{10.1103/PhysRevLett.92.062301}},
\href{http://www.arXiv.org/abs/nucl-ex/0310029}{\texttt{arXiv:nucl-ex/0310029}}.
%%CITATION = NUCL-EX/0310029;%%.

\bibitem{Poskanzer:1998yz}
\hrefCMSnoop {}{A.~M. Poskanzer and S.~A. Voloshin, ``Methods for analyzing
  anisotropic flow in relativistic nuclear collisions'',} \textit{ Phys. Rev.
  C} \textbf{ 58} (1998) 1671,
  \href{http://dx.doi.org/10.1103/PhysRevC.58.1671}{\doi{10.1103/PhysRevC.58.1671}},
\href{http://www.arXiv.org/abs/nucl-ex/9805001}{\texttt{arXiv:nucl-ex/9805001}}.
%%CITATION = NUCL-EX/9805001;%%.

\bibitem{Luzum:2012da}
\hrefCMSnoop {}{M.~Luzum and J.-Y. Ollitrault, ``Eliminating experimental bias
  in anisotropic-flow measurements of high-energy nuclear collisions'',}
  \textit{ Phys. Rev. C} \textbf{ 87} (2013) 044907,
  \href{http://dx.doi.org/10.1103/PhysRevC.87.044907}{\doi{10.1103/PhysRevC.87.044907}},
\href{http://www.arXiv.org/abs/1209.2323}{\texttt{arXiv:1209.2323}}.
%%CITATION = ARXIV:1209.2323;%%.

\bibitem{Adler:2002pu}
\hrefCMSnoop {}{{STAR} Collaboration, ``Elliptic flow from two and four
  particle correlations in {Au+Au} collisions at {$\sqrtsNN = 130\GeV$}'',}
  \textit{ Phys. Rev. C} \textbf{ 66} (2002) 034904,
  \href{http://dx.doi.org/10.1103/PhysRevC.66.034904}{\doi{10.1103/PhysRevC.66.034904}},
\href{http://www.arXiv.org/abs/nucl-ex/0206001}{\texttt{arXiv:nucl-ex/0206001}}.
%%CITATION = NUCL-EX/0206001;%%.

\bibitem{Chatrchyan:2014fea}
\hrefCMSnoop {}{{CMS Collaboration}, ``Description and performance of track and
  primary-vertex reconstruction with the {CMS} tracker'',} \textit{ JINST}
  \textbf{ 9} (2014) P10009,
  \href{http://dx.doi.org/10.1088/1748-0221/9/10/P10009}{\doi{10.1088/1748-0221/9/10/P10009}},
\href{http://www.arXiv.org/abs/1405.6569}{\texttt{arXiv:1405.6569}}.
%%CITATION = ARXIV:1405.6569;%%.

\bibitem{Chatrchyan:2008zzk}
\hrefCMSnoop {}{{CMS Collaboration}, ``The {CMS} experiment at the {CERN}
  {LHC}'',} \textit{ JINST} \textbf{ 3} (2008) S08004,
  \href{http://dx.doi.org/10.1088/1748-0221/3/08/S08004}{\doi{10.1088/1748-0221/3/08/S08004}}.

\bibitem{Agostinelli:2002hh}
\hrefCMSnoop {}{{GEANT4} Collaboration, ``{\GEANTfour}---a simulation
  toolkit'',} \textit{ Nucl. Instrum. Meth. A} \textbf{ 506} (2003) 250,
\href{http://dx.doi.org/10.1016/S0168-9002(03)01368-8}{\doi{10.1016/S0168-9002(03)01368-8}}.
%%CITATION = NUIMA,A506,250;%%.

\bibitem{Khachatryan:2016bia}
\hrefCMSnoop {}{{CMS Collaboration}, ``{The {CMS} trigger system}'',} \textit{
  JINST} \textbf{ 12} (2017) P01020,
  \href{http://dx.doi.org/10.1088/1748-0221/12/01/P01020}{\doi{10.1088/1748-0221/12/01/P01020}},
\href{http://www.arXiv.org/abs/1609.02366}{\texttt{arXiv:1609.02366}}.
%%CITATION = ARXIV:1609.02366;%%.

\bibitem{Khachatryan:2016odn}
\hrefCMSnoop {}{{CMS Collaboration}, ``Charged-particle nuclear modification
  factors in {PbPb} and {pPb} collisions at {$\sqrtsNN = 5.02\TeV$}'',}
  \textit{ JHEP} \textbf{ 04} (2017) 039,
  \href{http://dx.doi.org/10.1007/JHEP04(2017)039}{\doi{10.1007/JHEP04(2017)039}},
\href{http://www.arXiv.org/abs/1611.01664}{\texttt{arXiv:1611.01664}}.
%%CITATION = ARXIV:1611.01664;%%.

\bibitem{Lokhtin:2005px}
\hrefCMSnoop {}{I.~P. Lokhtin and A.~M. Snigirev, ``A model of jet quenching in
  ultrarelativistic heavy ion collisions and high-\pt hadron spectra at
  {RHIC}'',} \textit{ Eur. Phys. J. C} \textbf{ 45} (2006) 211,
  \href{http://dx.doi.org/10.1140/epjc/s2005-02426-3}{\doi{10.1140/epjc/s2005-02426-3}},
\href{http://www.arXiv.org/abs/hep-ph/0506189}{\texttt{arXiv:hep-ph/0506189}}.
%%CITATION = HEP-PH/0506189;%%.

\bibitem{Pierog:2013ria}
T.~Pierog\hrefCMSnoop {}{ {et~al.}, ``{EPOS LHC}: Test of collective
  hadronization with data measured at the {CERN Large Hadron Collider}'',}
  \textit{ Phys. Rev. C} \textbf{ 92} (2015) 034906,
  \href{http://dx.doi.org/10.1103/PhysRevC.92.034906}{\doi{10.1103/PhysRevC.92.034906}},
\href{http://www.arXiv.org/abs/1306.0121}{\texttt{arXiv:1306.0121}}.
%%CITATION = ARXIV:1306.0121;%%.

\bibitem{Pang:2014pxa}
L.-G. Pang\hrefCMSnoop {}{ {et~al.}, ``Longitudinal decorrelation of
  anisotropic flows in heavy-ion collisions at the {CERN Large Hadron
  Collider}'',} \textit{ Phys. Rev. C} \textbf{ 91} (2015) 044904,
  \href{http://dx.doi.org/10.1103/PhysRevC.91.044904}{\doi{10.1103/PhysRevC.91.044904}},
\href{http://www.arXiv.org/abs/1410.8690}{\texttt{arXiv:1410.8690}}.
%%CITATION = ARXIV:1410.8690;%%.

\bibitem{Acharya:2017zfg}
\hrefCMSnoop {}{{ALICE Collaboration}, ``Linear and non-linear flow modes in
  {Pb-Pb} collisions at {$\sqrtsNN=2.76\TeV$}'',} \textit{ Phys. Lett. B}
  \textbf{ 773} (2017) 68,
  \href{http://dx.doi.org/10.1016/j.physletb.2017.07.060}{\doi{10.1016/j.physletb.2017.07.060}},
\href{http://www.arXiv.org/abs/1705.04377}{\texttt{arXiv:1705.04377}}.
%%CITATION = ARXIV:1705.04377;%%.

\bibitem{Bilandzic:2013kga}
A.~Bilandzic\hrefCMSnoop {}{ {et~al.}, ``Generic framework for anisotropic flow
  analyses with multiparticle azimuthal correlations'',} \textit{ Phys. Rev. C}
  \textbf{ 89} (2014) 064904,
  \href{http://dx.doi.org/10.1103/PhysRevC.89.064904}{\doi{10.1103/PhysRevC.89.064904}},
\href{http://www.arXiv.org/abs/1312.3572}{\texttt{arXiv:1312.3572}}.
%%CITATION = ARXIV:1312.3572;%%.

\bibitem{Gardim:2011xv}
\hrefCMSnoop {}{F.~G. Gardim, F.~Grassi, M.~Luzum, and J.-Y. Ollitrault,
  ``Mapping the hydrodynamic response to the initial geometry in heavy-ion
  collisions'',} \textit{ Phys. Rev. C} \textbf{ 85} (2012) 024908,
  \href{http://dx.doi.org/10.1103/PhysRevC.85.024908}{\doi{10.1103/PhysRevC.85.024908}},
\href{http://www.arXiv.org/abs/1111.6538}{\texttt{arXiv:1111.6538}}.
%%CITATION = ARXIV:1111.6538;%%.

\bibitem{Yan:2015lwn}
\hrefCMSnoop {}{L.~Yan, S.~Pal, and J.-Y. Ollitrault, ``Nonlinear hydrodynamic
  response confronts {LHC} data'',} \textit{ Nucl. Phys. A} \textbf{ 956}
  (2016) 340,
  \href{http://dx.doi.org/10.1016/j.nuclphysa.2016.01.010}{\doi{10.1016/j.nuclphysa.2016.01.010}},
\href{http://www.arXiv.org/abs/1601.00040}{\texttt{arXiv:1601.00040}}.
%%CITATION = ARXIV:1601.00040;%%.

\bibitem{Lin:2004en}
Z.-W. Lin\hrefCMSnoop {}{ {et~al.}, ``A multi-phase transport model for
  relativistic heavy ion collisions'',} \textit{ Phys. Rev. C} \textbf{ 72}
  (2005) 064901,
  \href{http://dx.doi.org/10.1103/PhysRevC.72.064901}{\doi{10.1103/PhysRevC.72.064901}},
\href{http://www.arXiv.org/abs/nucl-th/0411110}{\texttt{arXiv:nucl-th/0411110}}.
%%CITATION = NUCL-TH/0411110;%%.

\bibitem{McDonald:2016vlt}
S.~McDonald\hrefCMSnoop {}{ {et~al.}, ``Hydrodynamic predictions for {Pb+Pb}
  collisions at 5.02 {\TeV}'',} \textit{ Phys. Rev. C} \textbf{ 95} (2017)
  064913,
  \href{http://dx.doi.org/10.1103/PhysRevC.95.064913}{\doi{10.1103/PhysRevC.95.064913}},
\href{http://www.arXiv.org/abs/1609.02958}{\texttt{arXiv:1609.02958}}.
%%CITATION = ARXIV:1609.02958;%%.

\end{thebibliography}\endgroup

\cleardoublepage \appendix\section{The CMS Collaboration \label{app:collab}}\begin{sloppypar}\hyphenpenalty=5000\widowpenalty=500\clubpenalty=5000\vskip\cmsinstskip
\textbf{Yerevan Physics Institute, Yerevan, Armenia}\\*[0pt]
A.M.~Sirunyan$^{\textrm{\dag}}$, A.~Tumasyan
\vskip\cmsinstskip
\textbf{Institut f\"{u}r Hochenergiephysik, Wien, Austria}\\*[0pt]
W.~Adam, F.~Ambrogi, T.~Bergauer, J.~Brandstetter, M.~Dragicevic, J.~Er\"{o}, A.~Escalante~Del~Valle, M.~Flechl, R.~Fr\"{u}hwirth\cmsAuthorMark{1}, M.~Jeitler\cmsAuthorMark{1}, N.~Krammer, I.~Kr\"{a}tschmer, D.~Liko, T.~Madlener, I.~Mikulec, N.~Rad, J.~Schieck\cmsAuthorMark{1}, R.~Sch\"{o}fbeck, M.~Spanring, D.~Spitzbart, W.~Waltenberger, C.-E.~Wulz\cmsAuthorMark{1}, M.~Zarucki
% ...
\vskip\cmsinstskip
\dag: Deceased\\
1:  Also at Vienna University of Technology, Vienna, Austria\\
% ...
\end{sloppypar}
%%% END EDITABLE REGION %%%
\end{document}